\documentclass[pra,twocolumn,showpacs,superscriptaddress,preprintnumbers,amsmath,amssymb,floatfix]{revtex4}
\usepackage{bm}
\usepackage{graphicx}
\usepackage{amsbsy}
\usepackage{amsmath}
\usepackage{amsfonts}
\usepackage{amsthm}
\usepackage{color}
\usepackage{mathrsfs}
\usepackage{graphicx}
\usepackage{dcolumn}
\usepackage{bm}

\newcommand{\bwt}{\begin{widetext}}
\newcommand{\ewt}{\end{widetext}}
\newcommand{\bea}{\begin{eqnarray}}
\newcommand{\eea}{\end{eqnarray}}

\begin{document}

\title{Quantum Computing on Multi-atomic Ensembles in Quantum Electrodynamics Cavity}

\author{Farid Ablayev}
\affiliation{Institute for Informatics of Tatarstan Academy of Sciences, 20 Mushtary, Kazan, 420012, Russia}
\affiliation{Kazan Federal University, Kremlevskaya 18, Kazan, 420008, Russia}.

\author{Sergey N. Andrianov}
\affiliation{Institute for Informatics of Tatarstan Academy of Sciences, 20 Mushtary, Kazan, 420012, Russia}


\author{Alexander Vasiliev}
\affiliation{Institute for Informatics of Tatarstan Academy of Sciences, 20 Mushtary, Kazan, 420012, Russia}
\affiliation{Kazan Federal University, Kremlevskaya 18, Kazan, 420008, Russia}.

\author{Sergey A. Moiseev}
\email{samoi@yandex.ru}

\affiliation{Institute for Informatics of Tatarstan Academy of Sciences, 20 Mushtary, Kazan, 420012, Russia}
\affiliation{Kazan Physical-Technical Institute of the Russian Academy of Sciences, 10/7 Sibirsky Trakt, Kazan, 420029, Russia}


\date{\today}

\begin{abstract}

We propose an effective realization of a complete set of elementary quantum
gates in the solid-state quantum computer based on the multi-atomic coherent
(MAC-) ensembles in the QED cavity. Here, we use the two-ensemble qubit
encoding and swapping-based operations that together provide implementation
of any encoded single-qubit operation by three elementary gates and the
encoded controlled-NOT operation is performed in a single step. This
approach simplifies a physical realization of universal quantum computing
and adds the immunity to a number of errors. We also demonstrate that the
proposed architecture of quantum computer satisfies DiVincenzo criteria.

\end{abstract}

\pacs{03.67.-a, 42.50.Ex}


\maketitle

\section{Introduction}
\label{sec:introduction}

During the last two decades different types of quantum computer and its
physical implementations have been considered \cite{Nielson2000,Nakahara2008, Kok2007, Ladd2010}, where single
natural or artificial atoms, ions, molecules etc., are used for encoding of
the qubits. Physical implementation of the quantum computing on many qubits
remains a main challenge in the first turn because of decoherence problems
causing intensive search for the novel experimental approaches. The
promising approach is that using natural atoms (ions, molecules,{\ldots})
with long coherence time.

Recently, a new physical realization of a quantum computer has been proposed
which uses multi-atomic coherent (MAC) ensembles for encoding of single
qubits \cite{Brion2007, Saffman2008}. MAC ensembles provide a huge enhancement of the effective
dipole moment that leads to a considerable acceleration of the quantum
processing rate. However, excess excited quantum states in the MAC- ensemble
should be blocked in order to realize an effective two-level system
providing thereby perfect encoding of the qubits. A dipole-dipole
interaction is intensively discussed for the blockade of the excess quantum
states \cite{Saffman2010}. However, the dipole blockade mechanism introduces the
decoherence due to a strong dependence of the dipole-dipole
interaction on a spatial distance between the interacting atoms. Recently,
another blockade mechanism based on using a light-shift imbalance in a Raman
transition has been proposed \cite{Shahriar2007} though its implementation is complicated because of additional quantum transitions arising with this imbalance.
We have also proposed a novel decoherence free blockade mechanism \cite{Moiseev2010, Andrianov2011, Moiseev2011}
 based on the collective interactions of the atoms in the QED cavity.
Rapid development of physics and technology of the microcavities \cite{Duan2004, Aoki2006, Majer2007}
 makes this blockade mechanism a quite promising though not very simple
for experimental realization.

By developing this approach, we demonstrate in this paper how logical single
and two-qubit gates can be  realized naturally in the quantum computer based
on the MAC- ensembles in the QED cavity using multi-qubit encoding
intensively discussed in
\cite{DiVincenzo1995,Deutsch1995, Boykin2000, DiVincenzo2000, Bacon2000, Kempe2001, Kempe2001b,
Kempe2002, Levy2002, Palma1996, Byrd2002}. Here, we explicitly show an encoded
universality for some set of these gates by using the 2-qubit encoding. This encoding
allows solving two major problems of solid-state quantum computing. First of
all, it eliminates the problems with implementation of single-qubit
operations. Additionally, the encoding forms a decoherence-free subspace
(DFS), which allows preventing a number of computational errors. Moreover,
this approach opens the ability of performing the controlled-NOT operation
in a single step. This operation is achieved by using an additional
nonlinear frequency shift of the atomic transitions due to Heisenberg-type
atom-atom interactions arising in the QED cavity. Finally we show that the
proposed implementation of the quantum gates set can be applied for the
construction of quantum computer satisfying DiVincenzo criteria \cite{DiVincenzo1998}.

\section{Swapping gates}

\begin{figure}
  \includegraphics[width=0.4\textwidth,height=0.25\textwidth]{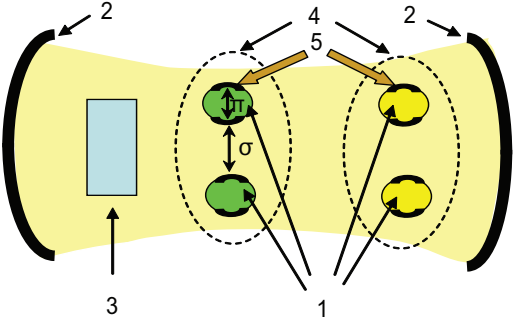}
  \caption
  {Scheme of a quantum computer based on multi-atomic ensembles (1) in
single mode QED cavity with mirrors (2). The QED-cavity is coupled to external flying photon qubits. Quantum
memory (3) is used for storage of the photon qubits. The qubits are transferred
to two pairs of processing nodes (green and yellow pairs) for realization of
single and two qubit gates. Each pair of nodes ((4) in dashed oval) is used for
encoding of a single qubit. Arrows (5) show microcavities, $\pi$ and $\sigma$ are the microcavity and cavity modes.}.
  \label{Figure1}
\end{figure}

We consider the realization of swapping gates controlled by a photon qubit.
We can launch signal photon qubits (denoted by $E_{in}$ and $E_{out})$ in
the QED cavity and take out the qubits in the free space through a
semitransparent mirror Fig. (\ref{Figure1}).
Many photon qubits can be efficiently
stored in quantum memory node also situated in the QED-cavity \cite{Moiseev2010b,Afzelius2010} and
then can be transferred to the processing nodes on demand \cite{Moiseev2011} for
implementation of single and two-qubit gates. For realization of controlled
swapping gate, we consider a logical qubit encoded on two MAC ensembles used
as processor nodes situated in two distant positions inside the common
QED-cavity. First, signal photon is transferred to one of the processor
nodes from the quantum memory or another logical qubit and the atomic
frequency in the processor nodes is tuned out of the cavity- mode frequency.
With that, mirrors are chosen to be dumb at new frequency of processing
nodes. The control photon is released from the node with control qubit and is
introduced into microcavity with the processing node via the
interaction with one MAC ensemble of logical qubit. Here, initially the control photon transfers to the MAC
ensemble due to the resonant interaction with additional atomic resonant
transitions when the frequency of control qubit node being detuned from the
resonant interaction. Then the control photon is released from the MAC
ensemble at frequency of the microcavity mode after tuning the frequency of
additional atomic transition to that of the microcavity. Afterwards, the
frequency of the additional atomic transition is detuned from the frequency
of the microcavity mode when control photon is already released. The
released control photon cannot be absorbed by MAC ensemble or escape from
the microcavity.

Let's introduce the following basis states of the atoms in two nodes 1 and
2: $\left| \psi \right\rangle _1 =\left| 1 \right\rangle _1 \left| 0
\right\rangle _2 $ and $\left| \psi \right\rangle _2 =\left| 0 \right\rangle
_1 \left| 1 \right\rangle _2 $, where in the case of sample that is small comparing
with wavelength of the cavity mode we use collective states $\left| 0 \right\rangle _m =\vert
0_1 ,0_2 ,...,0_{N_m } \rangle $, $\left| 1 \right\rangle _m =\sqrt {1/N}
\sum\limits_j^{N_m } {\left| {0_1 } \right\rangle \left| {0_2 }
\right\rangle ...\left| 1 \right\rangle _j ...\left| {0_{N_m } }
\right\rangle } $ for m-th node (N$_{m}$ is a number of atoms in m-th node)
By taking into account a presence of vacuum and single photon states in the
field modes $\pi _1 $ and $\pi _2 $ of 1-st and 2-nd
microcavities, we write the total wave function
as follows
\begin{align}
\label{eq1}
\left| \Psi \right\rangle =&\sum\limits_{n_{\pi _1 } =0}^1
\sum\limits_{n_{\pi _2 } =0}^1
\left| {n_{\pi _1 } }\right\rangle \left| {n_{\pi _2 } } \right\rangle
\nonumber  \\&
{\left\{ {c_1 \left( {n_{\pi _1 } ,n_{\pi _2
} } \right)\left| \psi \right\rangle _1 +c_2 \left( {n_{\pi _1 } ,n_{\pi _2
} } \right)\left| \psi \right\rangle _2 } \right\} }  ,
\end{align}
\noindent
where $n_{\pi _1 } $ and $n_{\pi _2 } $ are the numbers of photons
in the microcavity modes. 

Hamiltonian of system is written as $H=H_0 +H_1 $ where $H_0 =H_a +H_r $ is
the main Hamiltonian and $H_1 =H_{r-a} $ is perturbation Hamiltonian. Here, $H_a
=H_{a_1 } +H_{a_2 } $ is a Hamiltonian of the atoms in nodes 1 and 2 and $H_r
=H_\sigma +H_\pi $ is a Hamiltonian of photons in mode $\sigma $ of common QED
cavity, so that
$H_\pi =H_{\pi _1 } +H_{\pi _2 } .$ With that $H_{a_1 } =\hbar \omega _0
\sum\limits_{j_1 } {S_{j_1 }^z } $ and $H_{a_2 } =\hbar \omega _0
\sum\limits_{j_2 } {S_{j_2 }^z } $, where $\omega _0 $ is the working
frequency of atoms, $S_{j_1 }^z $ and $S_{j_2 }^z $are operators of
effective spin $\raise.5ex\hbox{$\scriptstyle 1$}\kern-.1em/
\kern-.15em\lower.25ex\hbox{$\scriptstyle 2$} $ for atoms in nodes 1 and 2;
$H_\sigma =\hbar \omega _{k_\sigma } a_{k_\sigma }^+ a_{k_\sigma } $, is a
Hamiltonian of the common QED-cavity field mode, $H_{\pi _1 } =\hbar \omega
_{k_{\pi _1 } } a_{k_{\pi _1 } }^+ a_{k_{\pi _1 } } $ and $H_{\pi _2 }
=\hbar \omega _{k_{\pi _2 } } a_{k_{\pi _2 } }^+ a_{k_{\pi _2 } } $ are the
Hamiltonians of the microcavity field modes, where $\omega _{k_\sigma } $
and $\omega _{k_{\pi _1 } } $, $\omega _{k_{\pi _2 } } $ are the frequencies
of photons with wave vectors $\vec {k}_\sigma $ and $\vec {k}_{\pi _1 } $,
$\vec {k}_{\pi _2 } $ of modes $\sigma $ and $\pi _{1}$, $\pi _{2}$,
$a_{k_\sigma }^+ $, $a_{k_{\pi _1 } }^+ $, $a_{k_{\pi _2 } }^+ $ and
$a_{k_\sigma } $, $a_{k_{\pi _1 } } $, $a_{k_{\pi _1 } } $ are creation and
annihilation operators of photons in corresponding modes.

For the interaction of photons with two atomic nodes $H_{r-a}
=H_{r-a}^{\left( 1 \right)} +H_{r-a}^{\left( 2 \right)} $, we have the
following expressions:
\begin{equation}
\label{eq2}
H_{r-a}^{\left( \alpha \right)} =\hbar \sum\limits_{j_\alpha }^{N_\alpha }
{\left( {g_{k_\sigma }^{(\alpha )} \exp \{i\vec {k}_\sigma \vec
{r}_{j_\alpha } \}S_{j_\alpha }^+ a_{k_\sigma } +H.C.} \right)} ,
\end{equation}
where $g_{k_\sigma }^{\left( \alpha \right)} $ is a constant of photon atom
interaction, $S_{j_2 }^+ $ is a raising operator for
effective spin $\raise.5ex\hbox{$\scriptstyle 1$}\kern-.1em/
\kern-.15em\lower.25ex\hbox{$\scriptstyle 2$} $ in two-level model, $\vec
{r}_{j_\alpha } $ are radius-vectors of atoms $j_\alpha $ in nodes $\alpha
=1,2$.

We are interested in nonresonant interaction between MAC ensembles and
the field modes. Here, one can use unitary transformation of
Hamiltonian $H_s =e^{-s}He^s$ leading in the second
perturbation order to:
\begin{equation}
\label{eq3}
H_s =H_0 +\frac{1}{2}\left[ {H_1 ,s} \right],
\end{equation}
when the relation $H_1 +\left[ {H_0 ,s} \right]=0$ holds. Using this
relation, we find $s=\sum\nolimits_{p=1}^4 {s_p } $, where
\begin{equation}
\label{eq4}
s_n =\hbar \sum\limits_{j_n }^{N_n } {\left( {\alpha _n g_{k_\sigma
}^{\left( n \right)} \exp \{i\vec {k}_\sigma \vec {r}_{j_n } \}S_{j_n }^+
a_{k_\sigma } -H.C.} \right)} ,
\end{equation}
\begin{equation}
\label{eq5}
s_{n+2} =\hbar \sum\limits_{j_n }^{N_n } {\left( {\beta _n g_{k_{\pi _n }
}^{\left( n \right)} \exp \{i\vec {k}_{\pi _n } \vec {r}_{j_n } \}S_{j_n }^+
a_{k_{\pi _n } } -H.C.} \right)} ,
\end{equation}
where $n=1,2$; $\hbar \alpha _n =1/\left( {\omega _{k_\sigma } -\omega _0 }
\right)=-1/\Delta _\sigma $, $\hbar \beta _n =1/(\omega _{k_{\pi _n } }
-\omega _0 )=-1/\Delta _{\pi _n } $.

Substituting expressions (\ref{eq4}) and (\ref{eq5}) into (\ref{eq3}), we get

\begin{align}
\label{eq6}
 H_s &= \hbar \omega _{k_\sigma } a_{k_\sigma }^+ a_{k_\sigma }
+\sum\limits_{m=1,2} {\hbar \omega _{k_{\pi _m } } a_{k_{\pi _m } }^+
a_{k_{\pi _m } } } +
\nonumber  \\&
\hbar \omega _1 \sum\limits_{j_1 =1}^{N_1 } {S_{j_1 }^z
} +\hbar \omega _2 \sum\limits_{j_2 =1}^{N_2 } {S_{j_2 }^z } +
\nonumber  \\&
2\hbar \sum\limits_{m=1,2} \{{\sum\limits_{j_m } [{\frac{\left| {g_{k_\sigma
}^{\left( m \right)} } \right|^2}{\Delta _m^{\left( \sigma \right)}
}a_{k_\sigma }^+ a_{k_\sigma } } } +
\frac{\left| {g_{k_{\pi _m } }^{\left( m \right)} }
\right|^2}{\Delta _m^{\left( {\pi _m } \right)} }a_{k_{\pi _m } }^+
a_{k_{\pi _m } } ]S_{j_m }^z   +
\nonumber  \\&
{\sum\limits_{i_m j_m } {\frac{\left|
{g_{k_\sigma }^{\left( m \right)} } \right|^2}{2\Delta _m^{\left( \sigma
\right)} }e^{i\vec {k}_\sigma \vec {r}_{i_m j_m } } } }
+ {\frac{\left| {g_{k_{\pi
_m } }^{\left( m \right)} } \right|^2}{2\Delta _m^{\left( {\pi _m } \right)}
}e^{i\vec {k}_{\pi _m } \vec {r}_{i_m j_m } }S_{i_m }^+ S_{j_m }^- } \} +
\nonumber  \\&
\frac{\hbar }{2}\left( {\frac{1}{\Delta _1^{\left( \sigma \right)}
}+\frac{1}{\Delta _2^{\left( \sigma \right)} }} \right)\sum\limits_{j_1 j_2
} {\left\{ {g_{k_\sigma }^{\left( 1 \right)} g_{k_\sigma }^{\left( 2
\right)\ast } e^{i\vec {k}_\sigma \vec {r}_{j_1 j_2 } }S_{j_1 }^+ S_{j_2 }^-
+h.c.} \right\}+}
\nonumber  \\&
\frac{\hbar }{2}\sum\limits_{m=1,2} {\sum\limits_{i_m }
\left({\frac{1}{\Delta _m^{\left( \sigma \right)} }+
\frac{1}{\Delta _m^{\left({\pi _m } \right)} }} \right) }
\nonumber  \\&
{\left\{ {g_{k_\sigma }^{\left( m \right)}
g_{k_{\pi _m } }^{\left( m \right)\ast } e^{i\left( {\vec {k}_\sigma -\vec
{k}_{\pi _m } } \right)\vec {r}_{i_m } } a_{k_\sigma } a_{k_{\pi
_m } }^+S_{i_m }^z +h.c.} \right\}}  .
 \end{align}

The first four terms in the Hamiltonian (\ref{eq6}) are unperturbed energy of photons
and atoms, the fifth and the sixth terms are atomic energy shifts due to
their interaction with photons, the seventh and the eighth terms are
intra-node interactions between atoms via virtual photons, the ninth term is
inter-node interaction between atoms via virtual photons and the tenth term
is inter-mode coupling. Hamiltonian (\ref{eq6}) describes non-resonant interaction
of atoms with the field modes which conserves the initial photon numbers of
the microcavity and common QED-cavity modes at $\Delta _m^{\left( \sigma
\right)} =-\Delta _m^{\left( {\pi _m } \right)} $ when the last term
vanishes due to the destructive interference between modes.

By assuming a possibility of single photon excitation only in $\pi _1 $ -
microcavity mode, we analyze the resonant interaction of two MAC-ensembles
via exchange of virtual photons of the common QED cavity mode. Using the
wave function (\ref{eq1}) for the case $n_{\pi _2 } =0$, we obtain the following
Schrodinger equation:
\begin{align}
\label{eq7}
& \frac{d\left| {\Psi \left( t \right)} \right\rangle }{dt}= -\frac{i}{\hbar
}H\left| {\Psi \left( t \right)} \right\rangle =
\nonumber  \\ &
i\sum\limits_{n_{\pi _1 }
=0}^1 {\left\{ [{\varpi _1 (n_{\pi _1 } )c_1 \left( {n_{\pi _1 } ,0}
\right) } \right.} -\sqrt {N_1 N_2 } \Omega
_\sigma c_2 \left( {n_{\pi _1 } ,0} \right)]\left| {\psi _1 } \right\rangle +
\nonumber  \\ &
[\varpi _2 (n_{\pi _1 } )c_2 \left( {n_{\pi _1 } ,0} \right)
-\left. {\sqrt {N_1 N_2 } \Omega _\sigma ^\ast c_1 \left(
{n_{\pi _1 } ,0} \right)]\left| {\psi _2 } \right\rangle } \right\} \left|
{n_{\pi _1 } } \right\rangle \left| 0 \right\rangle ,
 \end{align}

where
\begin{align}
\label{eq8}
\varpi _1 (n_{\pi _1 } )=&\left( {\frac{N_1 }{2}-1} \right)\left( {\omega _1
+2n_{\pi _1 } \Omega _1^{\left( {\pi _1 } \right)} } \right)+
\nonumber \\ &
\frac{N_2
}{2}\omega _2 -N_1 \left( {\Omega _1^{\left( \sigma \right)} +\Omega
_1^{\left( \pi \right)} } \right),
\end{align}
\begin{align}
\label{eq9}
\varpi _2 (n_{\pi _1 } )=&\frac{N_1 }{2}\left( {\omega _1 +2n_{\pi _1 }
\Omega _1^{\left( {\pi _1 } \right)} } \right)+
\nonumber \\ &
\left( {\frac{N_2 }{2}-1}
\right)\omega _2 -N_2 \left( {\Omega _2^{\left( \sigma \right)} +\Omega
_2^{\left( {\pi _2 } \right)} } \right),
\end{align}
where $\Omega _\sigma =\frac{g_{k_\sigma }^{\left( 1 \right)} g_{k_\sigma
}^{\left( 2 \right)\ast } }{2\hbar }\left( {\frac{1}{\Delta _1^{\left(
\sigma \right)} }+\frac{1}{\Delta _2^{\left( \sigma \right)} }} \right)$,
$\Omega _1^{\left( {\sigma ,\pi _1 } \right)} =\frac{\left| {g_{k_{\sigma
,\pi _1 } }^{\left( 1 \right)} } \right|^2}{\hbar \Delta _1^{\left( {\sigma
,\pi _1 } \right)} }$ and $\Omega _2^{\left( {\sigma ,\pi _2 } \right)}
=\frac{\left| {g_{k_{\sigma ,\pi _2 } }^{\left( 2 \right)} }
\right|^2}{\hbar \Delta _2^{\left( {\sigma ,\pi _2 } \right)} }$.

From (\ref{eq7}) we find the following equation for coefficients $c_1 \left( n
\right)\equiv c_1 \left( {n_{\pi _1 } ,0} \right)$, $c_2 \left( n
\right)\equiv c_2 \left( {n_{\pi _1 } ,0} \right)$ (where $n\equiv n_{\pi _1
} )$
\begin{equation}
\label{eq10}
\frac{d}{dt}c_1 \left( n \right)=i\varpi _1^{\left( n \right)} c_1 \left( n
\right)-i\sqrt {N_1 N_2 } \Omega _s c_2 \left( n \right),
\end{equation}
\begin{equation}
\label{eq11}
\frac{d}{dt}c_2 \left( n \right)=i\varpi _2^{\left( n \right)} c_2 \left( n
\right)-i\sqrt {N_1 N_2 } \Omega _s^\ast c_1 \left( n \right),
\end{equation}
Assuming the initial state with the excited first MAC-node $c_1 (n)=1$ and
$c_2 (n)=0$ we find the solution
\begin{equation}
\label{eq12}
c_1 (n)=e^{-i\varpi (n)t}\{\cos \left( {\kappa t} \right)+i\frac{\Delta
_{(n)} }{2\kappa _{(n)} }\sin \left( {\kappa _{(n)} t} \right)\},
\end{equation}
\begin{equation}
\label{eq13}
c_2 (n)=-ie^{-i\varpi (n)t}\frac{S}{\kappa _{(n)} }\sin \left( {\kappa
_{(n)} t} \right),
\end{equation}
where $\Delta _{(n)} /2=n\Omega _1^{\left( \pi \right)} $, $\kappa _{(n)}
=\sqrt {n^2\left( {\Omega _1^{\left( \pi \right)} } \right)^2+\left| S
\right|^2} $, $S=\sqrt {N_1 N_2 } \Omega _\sigma $, $\varpi
(n)=\frac{1}{2}\left( {\varpi _1^{\left( n \right)} +\varpi _2^{\left( n
\right)} } \right)$ and for convenience we have used the atomic parameters
satisfying the condition $\omega _2 -\omega _1 +N_2 \Omega _2^{\left( \sigma
\right)} -N_1 \Omega _1 =0$ (where $\Omega _1 \equiv \Omega _1^{\left(
\sigma \right)} +\Omega _1^{\left( \pi \right)} )$.

It can be seen in Eqs. (\ref{eq12}), (\ref{eq13}) that we have a strong blockade of
excitation transfer between the nodes at the presence of control photon
($n=1)$: $c_2 (n=1)\cong 0$ and $c_1 (n=1)\cong \exp \{-i[\varpi (n)+\Omega
_{_1 }^{\left( \pi \right)} ]t\}$ for sufficiently high quality factor of
the microcavity where $\Omega _{_1 }^{\left( \pi \right)} >>S$. Thus, there
is no swapping between the states $\left| {\psi _1 } \right\rangle $ and
$\left| {\psi _2 } \right\rangle $ of logical qubit in the presence of
control photon. While in the absence of control photon ($n=0)$, we have the
following oscillating solutions:
\begin{equation}
\label{eq14}
c_2 =-ie^{-i\varpi (n)t}\sin \left( {St} \right),
\end{equation}
\begin{equation}
\label{eq15}
c_1 =e^{-i\varpi (n)t}\cos \left( {St} \right),
\end{equation}
demonstrating periodical transfer of excitation between $\left| {\psi _1 }
\right\rangle $ and $\left| {\psi _2 } \right\rangle $ that is a realization
of Controlled-SWAP operation.

It is worth noting that the state $\left| \psi_1 \right\rangle $ is
transformed into the state $\left| \psi_2 \right\rangle $ at short time
intervals $t_{C-iSWAP} =\pi /(2N\Omega _S )$. By using of
interaction constants $g_\pi \sim 10^{10}Hz$ and $g_\sigma \sim 10^6Hz$, we get that
the number of atoms must be limited by sufficiently large quantity $N\le
10^4$, the time of energy transfer is $t_{C-iSWAP} \sim 10^{-8}\sec $ respectively.
So, we come to the realization of nano-dimensional swapping gates
controlled by the photon state. We call this gate
Controlled-iSWAP($\theta )$ gate because we have here the controlled rotation of
qubit on any desired angle $\theta =St$.

Another interesting case of the Controlled-iSWAP gate can be realized for
special quantum dynamics of Eqs. (\ref{eq12}) and (\ref{eq13}). Here, by using evolution
time $\tilde {t}=\pi /(2S)$ with interaction parameter $\vert \Omega
_1^{\left( \pi \right)} \vert =\sqrt 3 \vert S\vert $ one can provide a
perfect blockade of iSWAP operation due to $\kappa _{(\ref{eq1})} \tilde {t}=\pi $.
Eventual truth table for the iSWAP gate controlled be the two states of
photon ($\left| 0 \right\rangle $ and $\left| 1 \right\rangle )$ is the
following

\begin{table}[htbp]
\begin{center}
\begin{tabular}{|p{32pt}|p{30pt}|p{30pt}|p{64pt}|p{64pt}|}
\hline
&
$\left| \psi \right\rangle _1 \left| 0 \right\rangle $&
$\left| \psi \right\rangle _2 \left| 0 \right\rangle $&
$\left| \psi \right\rangle _1 \left| 1 \right\rangle $&
$\left| \psi \right\rangle _2 \left| 1 \right\rangle $ \\
\hline
$\left| \psi \right\rangle _1 \left| 0 \right\rangle $&
0&
$-i$&
0&
0 \\
\hline
$\left| \psi \right\rangle _2 \left| 0 \right\rangle $&
$-i$&
0&
0&
0 \\
\hline
$\left| \psi \right\rangle _1 \left| 1 \right\rangle $&
0&
0&
$e^{-i(\varpi (\ref{eq1})-\varpi (0))\tilde {t}}$&
0 \\
\hline
$\left| \psi \right\rangle _2 \left| 1 \right\rangle $&
0&
0&
0&
$e^{-i(\varpi (\ref{eq1})-\varpi (0))\tilde {t}}$ \\
\hline
\end{tabular}
\label{tab1}
\end{center}
\end{table}

Table 1. Truth table for the Controlled-iSWAP gate.
\noindent
This case imposes more mild relation between the quality factors of the
microcavities and common cavity. We note that it is possible also to realize the Controlled-iSWAP gate
where presence of a single photon will equalize the initially different node frequencies leading thereby
to the swapping process in an opposite manner to that in Table 1.

Summarizing, we note that realization of the Controlled-iSWAP gate provides
together with the iSWAP gate a Controlled-NOT gate by using logical
qubit encoding by two physical qubits. Below we discuss the main issues
related to a realization of complete set of gates for universal quantum
computations.

\section{ Universal quantum computing}

The principal scheme of a quantum computer with multiatomic ensembles in the
cavity is shown in Fig.1. Quantum memory is used for storage of many photon
qubits. The qubits are transferred to two pairs of processing nodes (blue
and green pairs) for realization of single- and two qubit gates. Each pair
of nodes (in dashed oval) is used for encoding of a single qubit.

According to DiVincenzo any quantum computer should have: 1. Scalability, 2.
Possibility of initialization, 3. Possibility of read-out, 4. Limited
decoherence, 5. Availability of a universal set of quantum gates. Let's
consider main DiVincenzo criteria \cite{DiVincenzo1998} for quantum computer in our
architecture.

1. Scalability. Our construction is scalable since the multi-qubit quantum memory and the processing nodes are
situated in common QED cavity \cite{Moiseev2011} and all nodes can be connected by swap
process providing the quantum operations over the large number of qubits stored in the quantum memory.

2. Initialization. Arbitrary initial quantum state of the system can be
reliably initialized and downloaded to the quantum memory node and qubits
can be transferred from quantum memory node to any processing node on
demand \cite{Moiseev2011}.

3. Read-out. A qubit state can be efficiently read out in a single-shot
fashion by detecting photon echo signals from the quantum memory incorporated in
the system \cite{Moiseev2010b}.

4. Decoherence. The system operates in decoherence free subspace and all
decoherence is connected with small inclinations from the model presented
here. Main sources of the decoherence are atomic phase relaxation ($\Gamma
)$ and cavity losses ($\gamma$). In this case we can generalize our result
by using Walls-Milburn input-out formalism \cite{Walls1994} for evaluation of the
fidelity for iSWAP operation $F=\vert \left\langle {\psi _{out} (\Gamma
,\gamma )} \mathrel{\left| {\vphantom {{\psi _{out} (\Gamma ,\gamma )} {\psi
_{out}^{ideal} }}} \right. \kern-\nulldelimiterspace} {\psi _{out}^{ideal} }
\right\rangle \vert ^2=\exp \{-2\Gamma t_{iSWAP} -\pi \gamma /2\Delta
\}\cosh^2(\pi \gamma /4\Delta )$ (where $t_{iSWAP} =\pi /(2N\Omega _\sigma ))$.
Thus realization of fault-tolerant quantum computing \cite{Knill2005} requires the
following values of the relaxation parameters $2\Gamma t_{iSWAP} +\pi \gamma
/2\Delta \le 10^{-4}$ that makes preferable using the spin transitions with
low decay constants and weak quantum noise $\Gamma $ \cite{Brown2011} in the QED cavities with high quality factor
\cite{Paik2011}. Similar requirements occur for iSWAP($\theta )$,
Controlled-iSWAP($\theta )$ and for other quantum gates in our architecture.

5. Universal set of quantum gates. The operation of a logical single qubit
(iSWAP) gate and two qubit gate (Control-iSWAP) was considered in the previous
chapter. Let's discuss the following property explicitly showing that in
encoding of qubits used here the two and three qubit operations
available in our physical model are sufficient to implement the standard
set, thus proving their encoded universality.

Property 2.1. The set of quantum gates {\{}Controlled-iSWAP; iSWAP($\theta )$;
PHASE($\phi )${\}}  is universal for the Hilbert subspace spanned by
encoded states $\vert 0_{L}>=\vert 01>, \vert 1_{L}> =\vert 10>$ (where PHASE($\phi )$- gate is realized by adjustable shifting of the atomic frequency $\Delta\omega$ in one of the two physical qubits  during the fixed time interval $t'= \phi/\Delta\omega$).

Proof. First of all we show the effect of our elementary quantum gates when
acting on pairs of nodes in basis states $\vert $01$>$, $\vert $10$>$ and
their linear combinations.

For instance, the iSWAP operation turns $\vert 01>$ into $|10>$ and
backwards, thus acting on a pair like the gate X (the NOT gate). In the
similar manner Controlled-SWAP actually implements the logical CNOT gate.

More formally:

CNOT$_{L}$ ($\alpha _{1}\vert $0$_{L}> \quad \vert $$0_L$$>+\alpha
_{2}\vert $0$_{L}>\vert $1$_{L}>+\alpha _{3}\vert
$1$_{L}>\vert $0$_{L}>+\alpha _{4}\vert $1$_{L}>\vert
$1$_{L}>)$ =

C(SWAP) ($\alpha _{1}\vert $01$>\vert $01$>+\alpha _{2}\vert
$01$>\vert $10$>+\alpha _{3}\vert $10$>\vert $01$>+\alpha
_{4}\vert $10$>\vert $10$>)$ =

$\alpha _{1}\vert $01$>\vert $01$>+\alpha _{2}\vert $01$>\vert
$10$>+\alpha _{3}\vert $10$>\vert $10$>+\alpha _{4}\vert
$10$>\vert $01$>$ =

$\alpha _{1}\vert $0$_{L}>\vert $0$_{L}>+\alpha _{2}\vert
$0$_{L}>\vert $1$_{L}>+\alpha _{3}\vert $1$_{L}>\vert
$1$_{L}>+\alpha _{4}\vert $1$_{L}>\vert $0$_{L}>$.

If we look at the matrix for the generalized iSWAP($\theta )$ gate, it's
middle part (responsible for transforming $\vert $01$>$ and $\vert $10$>$
basis states) is actually a rotation by the angle -$\theta $ about the x
axis of the Bloch sphere, i.e. iSWAP($\theta )$ corresponds to the following
operator:
\[
iSWAP\left( \theta \right)=\left( {{\begin{array}{*{20}c}
 1 \hfill & 0 \hfill & 0 \hfill & 0 \hfill \\
 0 \hfill & {\cos \frac{\theta }{2}} \hfill & {i\sin \frac{\theta }{2}}
\hfill & 0 \hfill \\
 0 \hfill & {i\sin \frac{\theta }{2}} \hfill & {\cos \frac{\theta }{2}}
\hfill & 0 \hfill \\
 0 \hfill & 0 \hfill & 0 \hfill & 1 \hfill \\
\end{array} }} \right)\to
\]
\[
R_x \left( \theta \right)=\left(
{{\begin{array}{*{20}c}
 {\cos \frac{\theta }{2}} \hfill & {i\sin \frac{\theta }{2}} \hfill \\
 {i\sin \frac{\theta }{2}} \hfill & {\cos \frac{\theta }{2}} \hfill \\
\end{array} }} \right).
\]
Similarly, PHASE($\theta )$ turns our composite qubit around the axis z (up
to the phase factor of $e^{i\varphi \mathord{\left/ {\vphantom {\varphi 2}}
\right. \kern-\nulldelimiterspace} 2})$:

\[
PHASE\left( \theta \right)=e^{i\varphi \mathord{\left/ {\vphantom {\varphi
2}} \right. \kern-\nulldelimiterspace} 2}\left( {{\begin{array}{*{20}c}
 {e^{-i\varphi \mathord{\left/ {\vphantom {\varphi 2}} \right.
\kern-\nulldelimiterspace} 2}} \hfill & 0 \hfill & 0 \hfill & 0 \hfill \\
 0 \hfill & {e^{-i\theta \mathord{\left/ {\vphantom {\theta 2}} \right.
\kern-\nulldelimiterspace} 2}} \hfill & 0 \hfill & 0 \hfill \\
 0 \hfill & 0 \hfill & {e^{i\theta \mathord{\left/ {\vphantom {\theta 2}}
\right. \kern-\nulldelimiterspace} 2}} \hfill & 0 \hfill \\
 0 \hfill & 0 \hfill & 0 \hfill & {e^{i\varphi \mathord{\left/ {\vphantom
{\varphi 2}} \right. \kern-\nulldelimiterspace} 2}} \hfill \\
\end{array} }} \right)\to
\]
\begin{equation}
\label{eq16}
R_z \left( \theta \right)=\left(
{{\begin{array}{*{20}c}
 {e^{-i\theta \mathord{\left/ {\vphantom {\theta 2}} \right.
\kern-\nulldelimiterspace} 2}} \hfill & \hfill \\
 \hfill & {e^{i\theta \mathord{\left/ {\vphantom {\theta 2}} \right.
\kern-\nulldelimiterspace} 2}} \hfill \\
\end{array} }} \right).
\end{equation}

Since an arbitrary rotation of a single qubit (and thus any single qubit
gate) can be decomposed into the product of three rotations about orthogonal
axes (say, R$_{x}$ and R$_{z})$, our basis allows to avoid using operations
of single processing nodes and thus blockage. All of the single qubit gates
are performed by the means two node operations.

For instance, the Hadamard transform can be implemented as follows:
\begin{equation}
\label{eq17}
H=e^{i\pi \mathord{\left/ {\vphantom {\pi 2}} \right.
\kern-\nulldelimiterspace} 2}R_z \left( {\frac{\pi }{2}} \right)R_x \left(
{\frac{\pi }{2}} \right)R_z \left( {\frac{\pi }{2}} \right).
\end{equation}
The other two single qubit gates S and T from the standard set up to the
phase factor are rotations about $z$ axis:
\begin{equation}
\label{eq18}
S=\left( {{\begin{array}{*{20}c}
 1 \hfill & 0 \hfill \\
 0 \hfill & i \hfill \\
\end{array} }} \right)=e^{i\pi \mathord{\left/ {\vphantom {\pi 4}} \right.
\kern-\nulldelimiterspace} 4}\left( {{\begin{array}{*{20}c}
 {e^{i\pi \mathord{\left/ {\vphantom {\pi 4}} \right.
\kern-\nulldelimiterspace} 4}} \hfill & 0 \hfill \\
 0 \hfill & {e^{-i\pi \mathord{\left/ {\vphantom {\pi 4}} \right.
\kern-\nulldelimiterspace} 4}} \hfill \\
\end{array} }} \right)=e^{i\pi \mathord{\left/ {\vphantom {\pi 4}} \right.
\kern-\nulldelimiterspace} 4}R_z \left( {\frac{\pi }{2}} \right),
\end{equation}
\begin{equation}
\label{eq19}
T=\left( {{\begin{array}{*{20}c}
 1 \hfill & 0 \hfill \\
 0 \hfill & {e^{i\pi \mathord{\left/ {\vphantom {\pi 4}} \right.
\kern-\nulldelimiterspace} 4}} \hfill \\
\end{array} }} \right)=e^{i\pi \mathord{\left/ {\vphantom {\pi 8}} \right.
\kern-\nulldelimiterspace} 8}\left( {{\begin{array}{*{20}c}
 {e^{-i\pi \mathord{\left/ {\vphantom {\pi 8}} \right.
\kern-\nulldelimiterspace} 8}} \hfill & 0 \hfill \\
 0 \hfill & {e^{i\pi \mathord{\left/ {\vphantom {\pi 8}} \right.
\kern-\nulldelimiterspace} 8}} \hfill \\
\end{array} }} \right)=e^{i\pi \mathord{\left/ {\vphantom {\pi 8}} \right.
\kern-\nulldelimiterspace} 8}R_z \left( {\frac{\pi }{4}} \right).
\end{equation}
Therefore, the set of gates {\{}Controlled-SWAP; iSWAP($\theta
)$;PHASE($\phi )${\}} allows to implement the standard set of quantum
gates, which proves it's encoded universality.

Note, that the universality of the proposed set of elementary gates rely on
the presence of ``continuous'' operations iSWAP($\theta )$ and PHASE($\theta
)$. This fact requires a higher precision of the hardware, but excludes the
approximation algorithms for implementing arbitrary single qubit operations
using the standard set of CNOT, H, S, and T. Conversely, we may restrict
ourselves with using only iSWAP($\pi $/2), PHASE($\pi $/2) and PHASE($\pi
$/4) (which is enough for implementing gates H, S, and T) and exploit
standard approximation schemes for arbitrary single qubit gates.

It is also well-known that the Controlled-SWAP (Fredkin) gate is universal
for classical computations, since it can be used to perform logical NOT,
AND, and FANOUT operations. Therefore, any classical computations can be
also performed by a quantum computer of the proposed architecture.

\section{Conclusion}

Summarizing, we have proposed an approach for constructing encoded universal
quantum computations on the multi-atomic coherent ensembles based on
swapping operations in QED cavity. The main ideas of the proposed quantum
computing scheme are encoding of logical qubits by two physical qubits and
using microcavities for implementing Controlled-iSWAP operations. Scalability is provided by using of multi-qubit quantum memory situated in one of the multi-atomic node. It allows
to implement explicitly any encoded single-qubit gate by 3 elementary
operations and to perform encoded controlled-NOT gate by a single
Controlled-SWAP operation on pairs of atomic ensembles. The physical
implementation of the basic gates is sufficiently robust and provides fast
single qubit operations based on multi-atomic ensembles. Detailed analysis
of these issues requires further investigation of physical limitations
determined by the experimental parameters of real physical systems.

The proposed approach considerably simplifies physical implementation of a
quantum computer on multi-atomic ensembles in the QED cavity at the price of
doubling the number of qubits for computation. Besides, it permits to avoid
the necessity of implementing blockade of excess states in the multi-atomic
ensembles. Note also, that using of two atomic ensembles for encoding of a
single qubit state will be convenient for the quantum computer interface
with the external quantum information carried by using photon polarization
qubits since the two polarization components of the photon qubit can be
coupled directly with the relevant pair of the atomic ensemble state.

\section{Acknowledgments}

Work was in part supported by the Russian Foundation for Basic Research
under the grants \# 09-01-97004, 10-02-01348, 11-07-00465.

\end{document}